\newif\ifsubmission
\submissiontrue

\newif\ifanonymous
\anonymousfalse

\documentclass[sigconf]{acmart}

\setcopyright{none}
\renewcommand\footnotetextcopyrightpermission[1]{}
\usepackage{etex}
\usepackage{endnotes}
\usepackage{booktabs}
\usepackage{graphicx}
\usepackage{amsmath}
\usepackage{multirow}
\usepackage{algorithmic}
\usepackage[ruled,vlined]{algorithm2e}
\usepackage{tikz}
\usepackage{cancel}
\usepackage{todonotes}
\usepackage{color}
\usepackage{subcaption}

\usepackage{enumitem}
\usepackage{listings}
\usepackage{numprint}
\usepackage{setspace}
\usepackage{placeins}
\usepackage{caption}
\usepackage{subcaption}
\usepackage{bigstrut}

\ifsubmission
  \newcommand{\TODO}[1]{}
  
  \newcommand{\REMOVE}[1]{}
\else
  \newcommand{\TODO}[1]{\textcolor{orange}{TODO\footnote{#1}}}
  
  \newcommand{\REMOVE}[1]{\textcolor{red}{#1}}
\fi

\newcommand{\sect}[1]{Section~\ref{#1}}
\newcommand{\fig}[1]{Figure~\ref{#1}}
\newcommand{\tab}[1]{Table~\ref{#1}}
\newcommand{\alg}[1]{Algorithm~\ref{#1}}

\newcommand{\sys}[1]{SkipGate}
\newcommand{\frwk}[1]{ARM2GC}

\newcommand{\BigO}[1]{\ensuremath{\mathcal{O}\bigl(#1\bigr)}}

\newcommand{\para}[1]{\vspace{3pt} \noindent \textbf{\textit{#1}}}

\npthousandsep{,}\npthousandthpartsep{}\npdecimalsign{.}
\graphicspath{{.}}
\definecolor{CommentColor}{RGB}{10,100,10}

\newcommand\Tstrut{\rule{0pt}{2.6ex}}         

\setlist[itemize]{noitemsep,nolistsep}
\setlist[itemize]{leftmargin=*}
\setlist[enumerate]{noitemsep,nolistsep}
\setlist[enumerate]{leftmargin=*}

\begin{document}
 \title{\huge \frwk{}: Succinct Garbled Processor for Secure Computation}

\author{Ebrahim M. Songhori}
\email{esonghori@google.com}
\affiliation{Google Inc.}
\author{M. Sadegh Riazi}
\email{mriazi@ucsd.edu}
\affiliation{UC San Diego}
\author{Siam U. Hussain}
\email{siamumar@ucsd.edu}
\affiliation{UC San Diego}
\author{Ahmad-Reza Sadeghi}
\email{ahmad.sadeghi@trust.tu-darmstadt.de}
\affiliation{TU Darmstadt}
\author{Farinaz Koushanfar}
\email{farinaz@ucsd.edu}
\affiliation{UC San Diego}

\begin{abstract}
{We present \frwk{}~\footnote{a short version of this paper to appear in Desgin Automation Conference (DAC) 2019}, a novel secure computation framework based on Yao's Garbled Circuit (GC) protocol and the ARM processor.
It allows users to develop privacy-preserving applications using standard high-level programming languages (e.g., C) and compile them using off-the-shelf ARM compilers (e.g., gcc-arm).
The main enabler of this framework is the introduction of \sys{}, an algorithm that dynamically omits the communication and encryption cost of the gates whose outputs are independent of the private data.
\sys{} greatly enhances the performance of \frwk{} by omitting costs of the gates associated with the instructions of the compiled binary, which is known by both parties involved in the computation.
Our evaluation on benchmark functions demonstrates that \frwk{} not only outperforms the current GC frameworks that support high-level languages, it also achieves efficiency comparable to the best prior solutions based on hardware description languages.
Moreover, in contrast to previous high-level frameworks with domain-specific languages and customized compilers, \frwk{} relies on standard ARM compiler which is rigorously verified and supports programs written in the standard syntax.}
\end{abstract}

\keywords{Privacy-Preserving Computation, Yao's Garbled Circuit, Secure Processor, ARM}



\begingroup
\mathchardef\UrlBreakPenalty=10000
\maketitle

\section{Introduction}\label{sec:intro}
Secure Function Evaluation (SFE) allows two or more parties to compute an arbitrary function on their respective inputs such that they learn the function's output without revealing their private data.
The first and one of the most powerful methods for two-party SFE is the Yao's Garbled Circuit (GC) protocol proposed by Andrew Yao in 1986~\cite{Yao86}.
Upon arrival, Yao's protocol immediately attracted significant attention from the cryptographic community and it has been the core enabler of many secure and privacy-preserving applications including but not limited to genomic data analysis~\cite{riazi2016genmatch}, text search~\cite{riazi2017prisearch}, and stable matching~\cite{riazi2017toward}.
The protocol requires representing the underlying function as a Boolean circuit.
It has been shown that the bottleneck of the GC protocol execution time is the communication between the two parties~\cite{gueron2015fast}. 
Therefore, the non-trivial challenge of utilizing GC is to generate the Boolean circuit such that its secure evaluation requires the minimum inter-party communication.

The challenge of the GC circuit optimization is partially addressed by TinyGarble~\cite{songhori2015tinygarble}.
This work shows that the GC-optimized circuit generation can be viewed as an atypical instance of the conventional logic synthesis task.
This approach outperforms previous methods for generating Boolean circuit using custom compilers or custom libraries~\cite{malkhi2004fairplay,buscher2017compiling, rastogi2014wysteria,demmler15aby,liu2015oblivm}.
In TinyGarble, however, the highest efficiency and scalability can only be achieved when the function is described in a Hardware Description Language (HDL), e.g., Verilog; while most users prefer to develop their applications in high-level programming languages, e.g., C.

In order to facilitate the deployment of the secure computation frameworks, many researchers have designed Domain Specific Languages (DSL) and/or custom designed compilers for secure computation~\cite{mood2016frigate,malkhi2004fairplay,kreuter2013pcf,franz2014cbmc,rastogi2014wysteria, buscher2017compiling}. These compilers enable users to write the program in a high-level language and compile them into a Boolean/Arithmetic circuit representation such that it can be evaluated by a secure computation protocol. Despite providing a user-friendly solution, the DSL and customized compilers exhibit many limitations compared to the standard high-level languages (e.g., C/C++). For example, they have specialized complex syntax, limited built-in data types, and certain rules on the programming style. Moreover, these compilers do not support many of the typical advanced code optimizations due to their customized design. Last but not the least, the recent analysis by Mood et al.~\cite{mood2016frigate} demonstrates that the current state-of-the-art high-level compilers for secure computation
crashed on programs that were compiled correctly or generated incorrect compiled programs in some cases. 

In this paper, we introduce \frwk{}, a novel 
\textit{garbled processor}
that supports developing privacy preserving applications using any (unmodified) high-level language 
and an off-the-shelf standard compiler (e.g., arm-gcc) without significant performance ramifications. In a garbled  processor~\cite{wang2015secure,songhori2016garbledcpu}, the underlying Boolean circuit is that of a general purpose processor. The compiled binary is loaded into the processor's instruction memory and the private data of the users is loaded into the data memory. 
Then, the circuit (processor) is garbled/evaluated through a GC back-end.
However, such a straightforward garbling approach results in a massive overhead compared to describing the program in HDLs or DSLs. 
For example, a single addition operation can be securely computed with a minimal number of gates when the task is described in an HDL~\cite{songhori2015tinygarble}. 
Performing the same task using a garbled processor requires garbling/evaluating all of the processor's components such as control path, register files, and the entire ALU.
Garbling these gates are not required to ensure privacy since they operate on the compiled binary of the function which is known to both parties. 

An earlier work by Wang et al.~\cite{wang2015secure} suggested a garbled processor based on the standard MIPS instructions. It incurred a high overhead compared with the recent DSL-based solutions. The reason for this inefficiency can be traced back to its {\it instruction-level} pruning of the processor circuit instead of {\it gate-level} optimization. (A more detailed description of this approach is discussed in \sect{sec:related}.)
\frwk{} benefits from the first \emph{dynamic fine-grained gate-level optimization} on the garbled processor such that only the gates associated with the private data incur garbling cost. The outputs of the gates associated with the instructions of the compiled binary of the function are computed locally by each party without communication or encryption. Moreover, the gates that do not contribute to the final output are dynamically skipped. This is enabled by the development of a novel algorithm called \sys{} that wraps around the GC protocol.
The algorithm dynamically computes the gate outputs that can be calculated without communication and marks the redundant gates for skipping.


The primary objective of \sys{} is to minimize the communication, the bottleneck of GC~\cite{gueron2015fast}, at the expense of a small increase in local computation.
Several secure computation compilers support somewhat similar approaches like {\it constant propagation} and {\it dead gate elimination}~\cite{paus2009practical,buscher2017compiling,mood2016frigate}. However, as we show in \sect{sec:skipgate} and \sect{sec:eval}, \sys{} is superior to these solutions since it operates at gate-level and does not require flattening the circuit for the entire computation; that is, it dynamically detects and removes gates that can be skipped from the garbling. 
Moreover, the static circuit simplification method~\cite{pinkas2009secure} that removes gates with constant inputs at compile time is not required by the \frwk{} framework, since the Boolean circuit of the ARM processor is generated by industrial circuit synthesis tools which take care of this task. 
Note that \sys{} avoids unnecessary garbling costs and is different from the cryptographic improvements of GC such as free-XOR~\cite{kolesnikov2008improved}, Row Reduction~\cite{naor1999privacy}, and Half Gate~\cite{zahur2015two} that reduce the garbling cost of an individual gate.
\sys{}'s operation is orthogonal to these methods; the underlying GC protocol in \frwk{} already benefits from these cryptographic improvements.



In contrast to the earlier custom high-level GC compilers, which employed ad-hoc verification techniques~\cite{kreuter2013pcf, franz2014cbmc, liu2015oblivm, zahur2015obliv}, \frwk{} inherits available ARM compilers. These compilers go through rigorous verification as they are used by a large community of programmers in different fields.
Therefore, it does not suffer from the reliability issues exposed by Frigate~\cite{mood2016frigate} in most of the state-of-the-art GC compilers. 
Moreover, it readily supports trivial simplifications such as $a = a\ \&\ a$, that is only supported by two of the most recent frameworks Frigate~\cite{mood2016frigate} and CBMC-GC~\cite{buscher2017compiling}.
Moreover, as our evaluation demonstrates, \frwk{} outperforms both these frameworks on the common benchmark functions. 


\vspace{1em}
\para{Contributions.}
\begin{itemize}
  \item We introduce \sys{}, the first algorithm that can dynamically optimize the sequential description of a garbled circuit to allow efficient secure evaluation of functions with publicly known inputs.
  \sys{} locally computes the output of the gates when it is independent of secret values.
  The algorithm also skips any gate which does not contribute to the final output. 

  \item We develop the \frwk{} framework based on the \sys{} algorithm and the ARM processor.
  In this framework, users can efficiently develop SFE applications in a high-level language like C/C++.
  It enables them to benefit from the available thoroughly verified compilers of ARM.
  We use the ARM architecture (without affecting the instruction set) to make it most effective for the GC protocol with \sys{}. 
  \item We provide extensive experimental results and show that \frwk{} is 156 times more efficient compared to prior garbled processors~\cite{wang2015secure, songhori2016garbledcpu}. 
  The \frwk{} framework demonstrates comparable performance to HDL synthesis approach of TinyGarble~\cite{songhori2015tinygarble}.
  \frwk{} also outperforms the state-of-the-art high-level GC compilers~\cite{buscher2017compiling, mood2016frigate} in terms of communication while utilizing unmodified programming languages and compilers.
\end{itemize}

\section{Preliminaries}
\subsection{Security Model}\label{ssec:security_model}
Consistent with the earlier relevant literature~\cite{holzer2012secure, rastogi2014wysteria,demmler15aby,liu2015oblivm,mood2016frigate}, we assume an \textit{honest-but-curious} adversary model where the participating parties follow the agreed upon protocol but may attempt to learn about the other parties' input from the information at hand~\cite{bellare2013efficient}.
This model can be generalized to more advanced adversary models that are typically addressed by multiple runs of the basic honest-but-curious model~\cite{lindell2007efficient, lindell2012secure}.

\subsection{Oblivious Transfer}
Oblivious Transfer (OT)~\cite{NaorP05} is a cryptographic protocol based on public key encryption executed between Alice (sender) and Bob (receiver) where Bob selects one of the messages provided by Alice without revealing his selection. Bob also does not learn anything about the unselected messages.
In an important special case of $1$-out-of-$2$ OT protocol ($\textrm{OT}^2_1$), Alice holds a pair of messages $(m_{0},\, m_{1})$; Bob holds a selection bit $b \in \{0, 1\}$ and obtains $m_{b}$ without revealing $b$ to Alice and learns nothing about $m_{1-b}$.

\subsection{Garbled Circuit}\label{ssec:preli_GC}
Yao's Garbled Circuit protocol~\cite{Yao86} allows two parties Alice (\textit{garbler}) and Bob (\textit{evaluator}) to jointly compute a function $c = f(a, b)$ on their private inputs ($a$ from Alice and $b$ from Bob) such that none of them reveal their inputs to each other.
In the end, one or both of them learn the output $c$.
The function $f$ is represented as a Boolean circuit consisting of 2-input gates.
For each wire $w$ in the circuit, Alice assigns two $k$-bit random keys, called \textit{labels}, $X_w^{0}$ and $X_w^{1}$ corresponding to 0 and 1 Boolean values respectively.
$k$ is the security parameter---typically $k=128$~\cite{bellare2013efficient}.
For each gate, Alice encrypts the output label in each row of the truth table with the corresponding input labels.
The resulting table containing the encrypted output labels is then randomly rearranged and called a \textit{garbled table}.
She sends the garbled tables of all gates along with the labels corresponding to her input values to Bob.
Bob obtains the labels corresponding to his input values obliviously through the OT protocol from Alice.
He uses these input labels to decrypt the garbled tables gate by gate.
In the end, Bob learns the labels for the final output wire and Alice has its mapping to 0 and 1 so that the actual value of the output can be determined.

The cost of communicating the garbled tables in the GC protocol is its performance bottleneck~\cite{gueron2015fast}.
Throughout the years, Yao's GC protocol has gone through a number of optimizations that reduce its communication cost.
We describe the most important optimizations here.
A significant optimization of the GC protocol is \textbf{\textit{free-XOR}}~\cite{kolesnikov2008improved} that removes the communication cost for XOR gates.
In this optimization, for any wire $w$, Alice only generates the label $X_w^{0}$ and computes the label corresponding to 1 as $X_w^{0}\oplus (R \parallel 1)$ where $\parallel$ represents bit concatenation and
$R$ is a global random ($k-1$)-bit value known only to Alice.
With this convention, the label for the output of an XOR gate with inputs $a$, $b$ and output $c$ can simply be computed as $X_{c} = X_{a} \oplus X_{b}$.
Thus, it does not need any encryption or transfer of garbled tables, meaning the XOR gate is \textit{free}.
As a result, the optimization goal for circuit generation is to minimize the number of non-XOR gates.

The \textbf{\textit{Row Reduction}}~\cite{naor1999privacy} lessens the communication cost of the AND gates by $25\%$ by generating the labels of the output wire as a function of the labels of the input wires and thus making one row of the garbled table all zeros.
The \textbf{\textit{Half Gate}} method~\cite{zahur2015two} utilizes both free-XOR and row reduction and reduces the cost of AND gates by an additional $25\%$.

Earlier GC protocols support only combinational circuit description of the logic functions.
Along with the use of logic synthesis for circuit generation, TinyGarble introduced the concept of the \textbf{\textit{Sequential Garbled Circuits}}~\cite{songhori2015tinygarble}.
Sequential circuits are cyclic graph representation of the Boolean circuits and allow for a compact representation of the functionality. Sequential circuits include memory elements (flip-flops) in addition to logic gates and run for multiple clock cycles.
In sequential GC, in each clock cycle, all the gates in the circuit are garbled/evaluated.
At the end of each cycle, the labels for the input wire of each flip-flop are simply transferred (copied) to its output wire to be used in the next cycle.
At the first clock cycle, the output wires of flip-flops are treated as (either Alice or Bob's) inputs depending on the function.

Utilizing sequential circuits drastically reduces the memory footprint during garbling and evaluation. For example, a 32-bit summation can be performed using a 1-bit full-adder circuit that outputs 1-bit of the result at each clock cycle. Another example of a sequential circuit is a {\it processor} that fetches the instruction, performs the corresponding computation, and stores the result.

\section{{\sys{}} Algorithm}\label{sec:skipgate}
\sys{} is a set of novel algorithms that automatically identifies gates that should be garbled given private and public inputs to the circuit. Any gate that can be evaluated based on the public values is {\it skipped} for garbling and is evaluated in plaintext instead. For example, consider a Multiplexer where both inputs are private and generated by some sub-circuits, whereas, the selection signal is public and known to both parties. 
In this scenario, one can skip the garbling of a sub-circuit that is not connected. 
However, standard garbling methodologies require the entire circuit to be garbled and to the best of our knowledge there is no systematic solution that can identify minimal set of gates that is necessary to be garbled.  

\sys{} is developed to complement the GC protocol for sequential circuits.
As explained in \sect{ssec:preli_GC}, GC allows secure computation of a function in the form $c = f(a, b)$. 
However, for a generic function,
in addition to the private inputs $a$ and $b$, there can be public inputs (known to both parties).
For example, in case of RSA, the encryption key is public. 
A more practical scenario is garbling a general purpose processor as we explain in \sect{sec:arm}.
In general, a processor will have two types of inputs: instruction and data, where the first input is known to both parties unless they want to keep the program private. 
If the GC framework does not distinguish between public and private inputs, garbling a processor will incur a massive cost for redundant garbling.
Previous work~\cite{songhori2015tinygarble, wang2015secure, songhori2016garbledcpu} proposed generating customized netlists for limited instruction sets. 
However, they fail to achieve the optimal optimization due to the coarse grain nature of their approach, i.e., instruction level as opposed to gate level.  

In \sys{}, we introduce a notion $p$ to incorporate the public inputs from both parties.
It allows secure evaluation of functions in the form of $c = f(a, b, p)$ where $p$ is the public input known to both parties and $a$ and $b$ are the private inputs.
The goal of \sys{} is to reduce the circuit of $c = f(a, b, p)$ into a simpler circuit $c = f_p(a,b)$ utilizing the knowledge of public input $p$.
Secure evaluation of $f_p(a,b)$ requires less number of garbled tables than that of $f(a, b, p)$ when using the standard GC protocol and treating $p$ as a private input.
\sys{} removes communication cost of garbling for a gate when its output can either be computed independently by Alice and Bob or has no effect on the final output.
In other words, it reduces the communication between the parties when it can be replaced by less costly local computation.
The cost reduction is especially significant in a garbled processor where the control path is public and independent of the private inputs.
Before presenting \sys{}, let us introduce the following notations and definitions.

In a classic Boolean circuit, each wire $w$ carries a value ($x_w\in\{0, 1\}$), whereas, in a garbled circuit, each wire carries a pair of labels ($X_w^{0}$ and $X_w^{1}$) on Alice's side and one label ($X_w \in \{X_w^{0}, X_w^{1}\}$) on Bob's.
If $X_w = X_w^{0}$, the actual Boolean value is 0 and if $X_w = X_w^{1}$, the value is 1.
This, in turn, means that the information is shared between the two parties.
In our scheme, we combine the notion of Boolean and garbled circuits.
Each wire either carries a Boolean value known to both parties independently (\textit{public} wire) or it carries a (pair of) label(s) (\textit{secret} wire).

\para{Illustrative Example}: Assume a sequential circuit that
has a 2-to-1 MUX whose inputs come from two sub-circuits f$_0$ and f$_1$ connecting to MUX inputs $0$ and $1$, respectively.
At a certain clock cycle, if the ``select wire'' of the MUX ($x$) is public, say equal to 1, both parties know that the gates in the sub-circuit f$_0$ do not need to be garbled/evaluated since they have no effect on the final output.
The gates in the MUX itself act as wires and pass the input $1$ (output of f$_1$) to the MUX output, thus they do not need to be garbled/evaluated in that clock cycle either.
However, in the conventional GC protocol where public wire $x$ is treated as a secret value, the entire circuit has to be garbled/evaluated.
In what follows, we explain how the \sys{} algorithm identifies such gates.


It is worth mentioning that in a sequential garbled circuit~\cite{songhori2015tinygarble}, the Boolean value of a wire can change at every clock cycle.
A wire may also switch between being secret and public from one clock cycle to another.
The \sys{} algorithm is executed once for every sequential cycle.
\sys{}'s decision on each gate
depends on the status of the gate's inputs (public or secret) on that cycle.

\subsection{Gate Categories}\label{ssec:skipgate_example}
The \sys{} algorithm classifies the gates into four categories in terms of the parties' knowledge about the inputs of a given gate:

\begin{enumerate}[label=\roman*]
  \item \textit{Gate with two public inputs:}
    In this case, the output is public and can be computed locally by each party.
  \item \textit{Gate with one public input:}
  	Depending on the gate type, the output becomes either public or secret.
  	For example, for an AND gate with public value 0 at one input, the output becomes 0.
  	This means that if the gate's secret input is not connected to any other gate, the gate generating the secret wire can be skipped for garbling/evaluation.
  	If the public input is 1, then the AND gate acts as a wire and the output wire carries the label of the secret input.
  \item \textit{Gate with secret inputs that have identical (or inverted) labels:} This indicates that the two secret inputs have identical (or inverted) Boolean values. Depending on the gate type, the output becomes either public or secret. For example, the output of an XOR gate with two inverted inputs (either secret or public) is always 1 (public). Similar to Category ii, the gate generating the inputs, if not connected to any other gate, can be skipped for garbling/evaluation.
  \item \textit{Gate with unrelated secret inputs:}
  	The output is always secret.
  	The gate has to be garbled/evaluated conventionally according to the GC protocol. However, if its output does not have any effect on the circuit output, the gate is skipped, i.e., the corresponding garbled table is not transferred.
\end{enumerate}

\subsection{Algorithm}\label{ssec:skipgate_algorithm}
\alg{alg:alice} and \alg{alg:bob} show the \sys{} algorithm for Alice and Bob sides, respectively.
Lines 2-5 of \alg{alg:alice} and Lines 2-4 of \alg{alg:bob} are similar to the GC protocol label generation and transfer for both sides.
The \sys{} algorithm has two main phases:
In Phase 1, the outputs of the gates with public input(s) (Categories i-ii) are computed.
In Phase 2, the gates with private inputs (Categories iii-iv) are garbled/evaluated.
For each round of sequential cycle, Alice executes Phase 1 and 2 of \sys{} and sends the generated garbled tables to Bob.
Bob receives the tables and executes two phases in order to evaluate the gates.
However, this does not affect the parallelism of the operation. When Bob is evaluating the gates in cycle $c$, Alice is garbling the gates for cycle $c+1$.
In Line 14 of \alg{alg:alice} and Line 13 of \alg{alg:bob}, the labels associated with the input of flip-flops are copied to their output for the next cycle~\cite{songhori2015tinygarble}.
Similar to conventional GC, at the end of the protocol, Alice learns pairs of labels for each output wire and Bob has one of the labels; they share this information to learn the output $c$.
For example, in the case where Alice intends to learn the final output, she receives Bob's output label and together with her output labels finds the real output value (Line 16-17 of \alg{alg:alice} and Line 16 of \alg{alg:bob}).

\begin{algorithm}[t]
\caption{\sys{}, Alice's side.}\label{alg:alice}
\textbf{Inputs:} Sequential circuit of $c=f(a,b,p)$, Alice's input $a$, public inputs $p$, number of clock cycles $cc$.\\
\textbf{Outputs:} Output $c$.\\
\begin{algorithmic}[1]
\STATE{\bf{\sys{}\_Alice\ (circuit, a, p, cc)}:}
\STATE{generate random labels $\rightarrow (X^0_A,X^1_A,X^0_B,X^1_B)$}
\STATE{send Alice labels $\in \{X^0_A, X^1_A\}$ based on her input $a$}
\STATE{send Bob labels $(X^0_B, X^1_B)$ through OT}
\STATE{set wires corresponding to $a$ and $b$ as private}
\STATE{set wires corresponding to $p$ as public}
\FOR{cid \textit{from} $1$ \textit{to} $cc$}
  \STATE{initialize labels' fanout}
  \STATE{\textit{// \alg{alg:phase1}, process gate categories i-ii}}
  \STATE{perform Phase 1}
  \STATE{\textit{// \alg{alg:phase2_alice}, process gate categories iii-iv}}
  \STATE{perform Alice Phase 2 $\rightarrow$ garbled tables}
  \STATE{send garbled tables}
  \STATE{copy flip flops labels}
\ENDFOR
\STATE{receive Bob output labels $\rightarrow$ $X_C$}
\STATE{compute output value based on output labels $(X^0_C,X^1_C)$ and received labels $(X_C)$  $\rightarrow$ $c$}
\end{algorithmic}
\end{algorithm}

\begin{algorithm}[t]
\caption{\sys{}, Bob's side.}\label{alg:bob}
\textbf{Inputs:} Sequential circuit of $c=f(a,b,p)$, \\Bob's input $b$, public input $p$, number of clock cycles $cc$.\\
\textbf{Outputs:} Output labels $X_C$.\\
\begin{algorithmic}[1]
\STATE{\bf{\sys{}\_Bob(circuit, b, p, cc)}:}
\STATE{receive Alice's labels $\rightarrow$ $X_A$}
\STATE{receive Bob labels $\rightarrow$ $X_B$ through OT}
\STATE{set wires corresponding to $a$ and $b$ as private}
\STATE{set wires corresponding to $p$ as public}
\FOR{cid \textit{from} $1$ \textit{to} $cc$}
  \STATE{initialize labels' fanout}
  \STATE{\textit{// \alg{alg:phase1}, process gate categories i-ii}}
  \STATE{perform Phase 1}
  \STATE{receive garbled tables}
  \STATE{\textit{// \alg{alg:phase2_bob}, process gate categories iii-iv}}
  \STATE{perform Bob Phase 2}
  \STATE{copy flip flops labels}
\ENDFOR
\STATE{compute circuit output labels $\rightarrow$ $X_C$}
\STATE{send output labels ($X_C$)}
\end{algorithmic}
\end{algorithm}

\begin{algorithm}[ht]
\caption{Phase 1 in \sys{} for both Alice and Bob sides.}\label{alg:phase1}
\begin{algorithmic}[1]
\STATE{\bf{\sys{}\_phase1()}:}
\FOR{g \textit{in} circuit}
	\IF{both inputs of g are public}
		\STATE{\textit{//Category i}}
		\STATE{compute output of g based on its type\\ and inputs}
		\STATE{set g label fanout to $0$}
	\ELSIF{one of the g inputs is public}
		\STATE{\textit{//Category ii}}
		\STATE{compute output of g based on its type, private, and public inputs}
		\IF{output of g is public}
			\STATE{set g label fanout to $1$ \textit{// will become zero in recursive\_reduction()}}
			\STATE{recursive\_reduction(g)}
		\ENDIF
	\ENDIF
\ENDFOR
\end{algorithmic}
\end{algorithm}

In \sys{}, an integer called \texttt{label\_fanout} is associated with each gate and indicates the number of times the gate's output label is used (either as a circuit's output or an input to other gates).
At the beginning of each cycle (Line 8 of \alg{alg:alice} and Line 7 of \alg{alg:bob}), the \texttt{label\_fanout} is set to the gate fanout in the circuit\footnote{\textit{Fanout} of a gate, borrowed from hardware design, is the number of subsequent gates (and circuit outputs) dependent on the gate's output.}.
\texttt{label\_fanout} of a gate may decrease, e.g., a gate whose output is connected to an AND gate with 0 at the other input (Category ii).
If \texttt{label\_fanout} reaches 0, it means that gate's output label does not have any effect on the rest of the circuit and final output.
The gates with \texttt{label\_fanout} $=0$ are subsequently marked for skipping, which in turn decreases the \texttt{label\_fanout} of the gates connected to the input of the marked gates. Note that this step is {\it recursive} in nature, i.e., when decreasing a gate's \texttt{label\_fanout}, it might reach zero and subsequently call the gates who are providing the input to this gate and so on (see \fig{fig:phaseOneRecursive}). 
Finally, the gates in Category iv that have not been marked for skipping are garbled/evaluated.

\alg{alg:phase1} illustrates the Phase 1 of \sys{} in which Alice and Bob find and compute the gates that belong to Categories i-ii.
\texttt{label\_fanout} of the gates in Category i are set to zero.
For gates in Category ii, if the output becomes public, \sys{} decreases the \texttt{label\_fanout} of the secret input's originating gate recursively by invoking \texttt{recursive\_reduction} (\alg{alg:skipgate_reduction}).
\fig{fig:phaseOneExample} shows four different examples in Phase 1.
Bob does not receive any information from Alice about the gates in Category i-ii because he can locally evaluate Phase 1 just like Alice.
An alternative approach is that Alice sends the result of Phase 1 to Bob.
This approach has two main disadvantages:
First, it makes the protocol altered if one wants to enhance the security of the protocol to be secure against malicious adversaries~\cite{lindell2007efficient}.
Second, it increases the communication overhead which is the bottleneck of the GC protocol~\cite{gueron2015fast}.

\begin{figure}[b]
    \centering
    \includegraphics[width = 0.45\columnwidth]{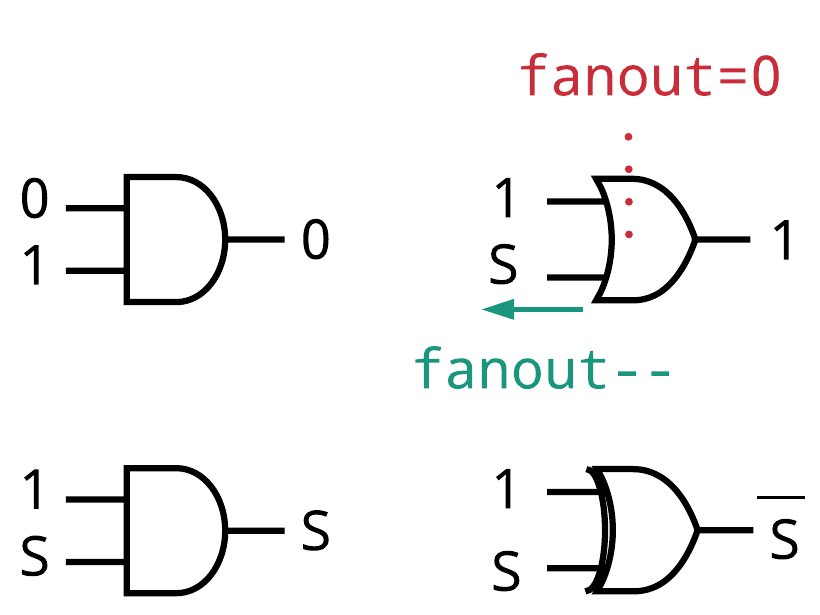}
    \caption{Four examples in Phase 1 where gates are replaced by zero, one, wire, or inverter.
    The top-left gate is in Category i and the rest are in Category ii.
    \texttt{label\_fanout} is set to zero for the skipped gate.}
    \label{fig:phaseOneExample}
\end{figure}

\begin{algorithm}[t]
\caption{Phase 2 in \sys{}, Alice's side.}\label{alg:phase2_alice}
\textbf{Output}: list of garbled tables.\\
\begin{algorithmic}[1]
\STATE{\bf{\sys{}.phase2\_Alice()}:}
\FOR{g \textit{in} circuit where label\_fanout $> 0$}
	\IF{g's input labels are equal or inverted}
		\STATE{\textit{//Category iii}}
		\STATE{compute g's output based on its type}
		\IF{g's output label is public}
			\STATE{set g's label\_fanout to $1$ \textit{// will become zero in recursive\_reduction()}}
			\STATE{recursive\_reduction(g)}
		\ENDIF
	\ELSE \STATE{\textit{//Category iv}}
    \STATE{garble g \textit{// table = null for XOR gates}}
    \IF{g is non-XOR}
      \STATE{add garbled table to the list}
    \ENDIF
	\ENDIF
\ENDFOR
\STATE{remove garbled tables where gates's fanout is $0$}
\end{algorithmic}
\end{algorithm}

\alg{alg:phase2_alice} shows the Phase 2 of \sys{} for Alice's side in which she performs the same task for Category iii.
She then generates garbled tables for gates with non-zero \texttt{label\_fanout} in Category iv.
\fig{fig:phaseTwoExample} shows four different examples in this phase.
By the end of Phase 2, due to the recursive nature of the fanout reduction, \texttt{label\_fanout} of some gates that have already been garbled may become 0.
In Line 18 of \alg{alg:phase2_alice}, Alice filters the garbled tables that have non-zero \texttt{label\_fanout} to be sent to Bob.

\begin{figure}[b]
    \centering
    \includegraphics[width = 0.62\columnwidth]{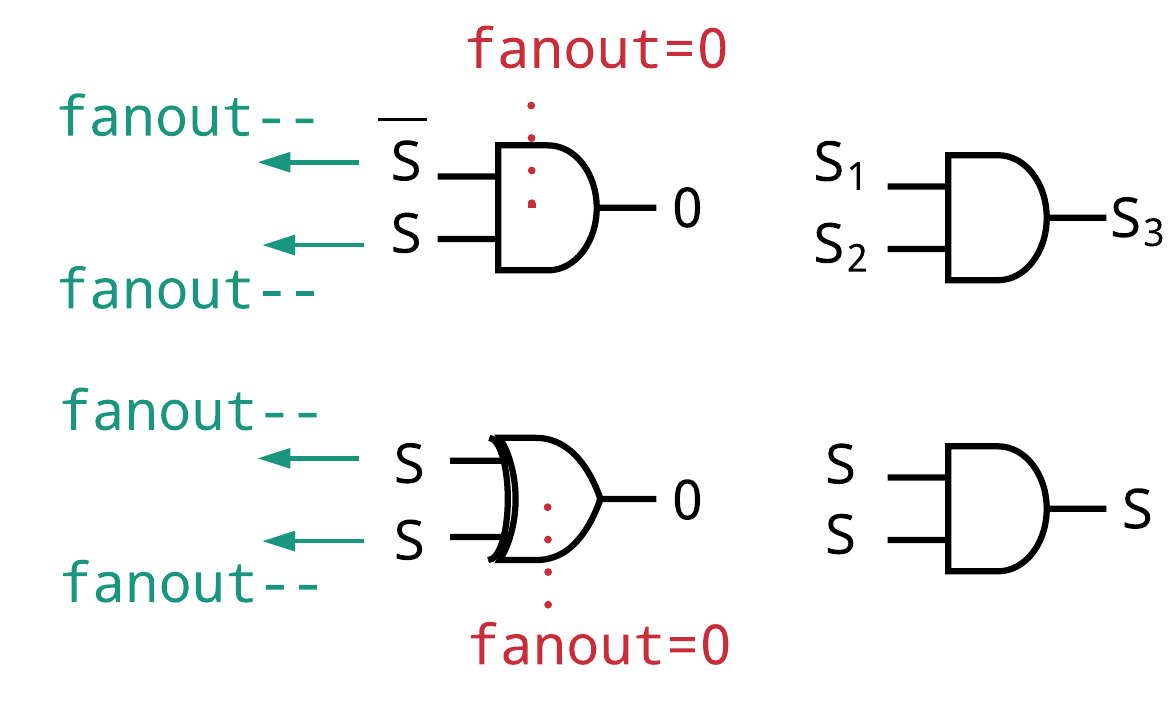}
    \caption{Four examples of replacing and computing gates in Phase 2. The top-right example is in Category iv, and the rest are in Category iii. \texttt{label\_fanout} is set to zero for the skipped gates. }
    \label{fig:phaseTwoExample}
\end{figure}

\begin{algorithm}[t]
\caption{Phase 2 in \sys{}, Bob's side.}\label{alg:phase2_bob}
\textbf{Input}: list of garbled tables.\\
\begin{algorithmic}[1]
\STATE{\bf{\sys{}.phase2\_Bob(garbled\_tables)}:}
\FOR{g \textit{in} circuit where label\_fanout $> 0$}
	\IF{g's input labels are equal or inverted}
		\STATE{\textit{//Category iii}}
		\STATE{compute g's output based on its type}
		\IF{g's output label is public}
			\STATE{set g's label\_fanout to $1$ \textit{// will become zero in recursive\_reduction()}}
			\STATE{recursive\_reduction(g)}
		\ENDIF
	\ELSE \STATE{\textit{//Category iv}}
    \IF{g is an XOR gate}
     \STATE{compute output label based on input labels}
    \ELSIF{g is top of the garbled tables list}
      \STATE{remove the garbled table from the list $\rightarrow$ gt}
      \STATE{compute output label of g based on its type, input labels, and gt}
    \ELSE
      \STATE{assign g's output label to a unique random binary string}
    \ENDIF
	\ENDIF
\ENDFOR
\end{algorithmic}
\end{algorithm}

\alg{alg:phase2_bob} shows the Phase 2 for Bob's side.
Bob evaluates the gates that belong to Category iii and iv.
In Line 18 of \alg{alg:phase2_bob}, Bob generates and assigns new unique labels for gates that were filtered by Alice.
Bob knows that the \texttt{label\_fanout} of these gates will eventually become $0$.
Therefore, he produces new labels for them only to keep track of these secret variables that are used to compute the output of the gates in Category iii.
He can generate these labels randomly or use a monotonic counter that increases by one for each newly generated label.
To distinguish valid GC labels from his generated labels, he keeps a single bit flag along with each label that indicates the label is generated by him and is not valid for the GC evaluation.

\begin{algorithm}
\caption{Recursive Fanout Reduction of \sys{}.}\label{alg:skipgate_reduction}
\textbf{Inputs:} Gate \texttt{g} (where the reduction starts).\\
\begin{algorithmic}[1]
\STATE{\bf{\sys{}.recursive\_reduction(g)}:}
\IF{g's label\_fanout is $0$}
	\STATE{return}
\ENDIF
\STATE{g's label\_fanout $=$ label\_fanout - $1$}
\IF{label\_fanout is $0$}
	\IF{g's first input is secret}
		\STATE{recursive\_reduction(first input of g)}
	\ENDIF
	\IF{g's second input is secret}
		\STATE{recursive\_reduction(second input of g)}
	\ENDIF
\ENDIF
\end{algorithmic}
\end{algorithm}

\alg{alg:skipgate_reduction} illustrates the pseudo-code for the recursive fanout reduction.
It receives the circuit and a gate inside the circuit.
It first decreases the \texttt{label\_fanout} of the given gate.
If the \texttt{label\_fanout} becomes 0, it recursively calls the function with the gates that generate the corresponding secret input(s).
This process is illustrated on an example circuit in \fig{fig:phaseOneRecursive}.

\subsection{Identification of Identical and Inverted Labels}
According to the GC protocol, Bob only has one label $X_w$ for each secret wire $w$.
Due to free-XOR~\cite{kolesnikov2008improved}, he does not need to modify the label when he evaluates a NOT gate because the labels corresponding to $0$ and $1$ are inverted by Alice during the garbling process, flipping the secret value of $w$ accordingly.
This, in turn, means that Bob cannot tell apart an identical and inverted secret value based on the label alone.
However, it is still possible for Bob to keep track of the flips by storing one bit along with the label.
After evaluating a NOT gate, he simply flips the bit.
This extra bit helps him to differentiate between identical and inverted secret values which are crucial during Phase 2.

\begin{figure}[hb]
    \centering
    \includegraphics[width=\columnwidth]{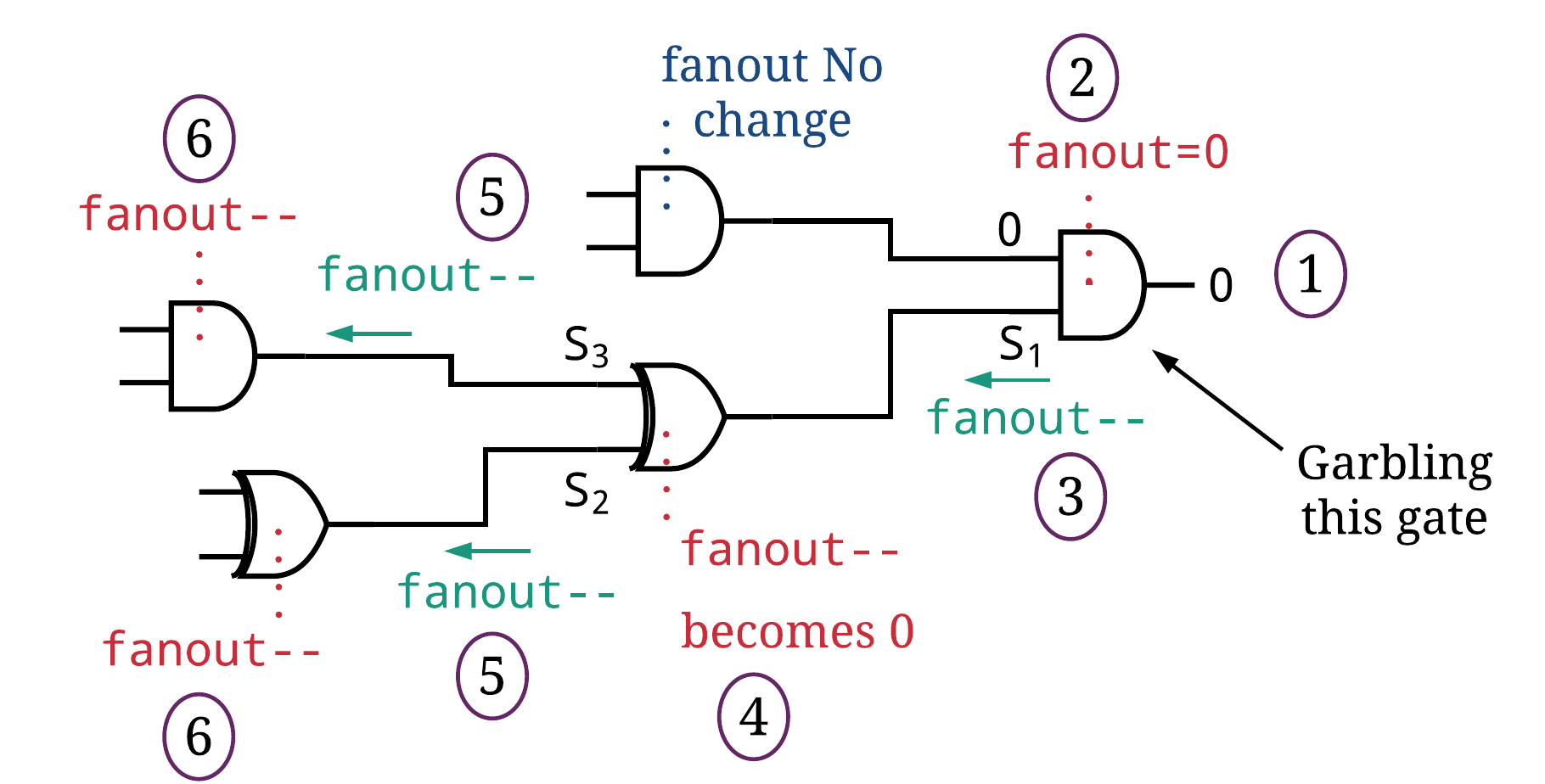}
    \caption{Recursive reduction of \texttt{label\_fanout} to skip unnecessary gates (\alg{alg:skipgate_reduction}).}
    \label{fig:phaseOneRecursive}
\end{figure}
%

\subsection{Computational Complexity}\label{ssec:skipgate_complexity}
The \sys{} algorithm decreases the communication cost, which is the bottleneck of GC, at the expense of increasing the local computations.
The conventional GC protocol has a linear computational complexity in terms of the number of gates in the circuit for each sequential cycle.
We show that, despite its recursive appearance, the \sys{} algorithm does not increase the computation complexity of the GC protocol.
All parts of the \sys{} algorithm, except \texttt{recursive\_reduction} (\alg{alg:skipgate_reduction}), is executed once per gate, thus they incur a complexity similar to the classic GC protocol.
The only procedure that can potentially increase the computation complexity is \texttt{recursive\_reduction} function whose number of invocations depends on the underlying circuit and whether input wires are secret or public.
To find the complexity of \sys{}, we present an upper bound on the number of invocations of \texttt{recursive\_reduction} function.

The termination condition in \texttt{recursive\_reduction} is the fanout reaching zero (Lines 2 of \alg{alg:skipgate_reduction}).
Thus, the worst case scenario is when the function reduces the fanout of all the gates to zero.
In this case, the number of execution of the fanout decrement (Line 5) should be at most the sum of all the initialized fanouts.
\texttt{label\_fanout} is initialized with the gate fanout in the circuit.
The upper bound on the sum of fanouts of all the gates in the circuit is $$F = \sum_{i=1}^{n} g[i].fanout \le 2n - m + q,$$ where $n$ is the number of gates, $q$, and $m$ are the number of circuit output and inputs, respectively.
Each gate has two inputs, as required by the GC protocol, 
and each input creates a fanout in previous gates unless it is a circuit input.
Also, each output wire incurs the fanout of one.
Both $q$ and $m$ are typically less than or at most in the order of $n$.
Thus, $F$ and subsequently the number of invocation of \texttt{recursive\_reduction} function are $\BigO{n}$.
This shows that \sys{} does not increase the overall linear computational complexity of the GC protocol.

\subsection{Correctness and Security Proof}
\para{Correctness}: Given the correctness of Yao's GC protocol, we show that GC protocol with \sys{} is also correct.
In \sys{}, the topology of the circuit is not changed, thus the dependencies of the values remain the same.
Therefore, we only prove that processing a single gate remains correct in \sys{}. 

The operations for gates in Category i are merely based on the Boolean operation of the gates and are clearly correct.
For gates in Categories ii-iii, the secret input is considered as an {\it unknown} variable.
Either the label at the secret input of the gate is passed to its output or the output is set to a public value.
Therefore, the functionality of the gate is not changed.
Gates in Category iv with non-zero \texttt{label\_fanout} are garbled/evaluated according to the GC protocol.
For the rest of the gates in Category iv, \texttt{label\_fanout} $=0$ indicates that their secret output does not have any effect on the final output of the circuit.
Therefore, they can be safely skipped.
As such, we conclude that the \sys{} algorithm with the GC protocol results in a logically correct output.

\para{Security}: The GC protocol is proved to be secure under honest-but-curious adversary model for any two-input Boolean function $f(a, b)$ where $a$ and $b$ are private inputs from Alice and Bob, respectively~\cite{lindell2009proof, bellare2013efficient}.
In this work, we extend the function to the form of $f(a, b, p)$ to include a public input $p$ that is known to both parties.
The \sys{} algorithm reduces the Boolean circuit of $f(a, b, p)$ to a two-input circuit of $f_p(a, b)$ where, for a given $p$, $f_p(a, b) = f(a, b, p)$ for any $a$ and $b$.
$f_p(a, b)$ consists of the gates in Category iv with non-zero \texttt{label\_fanout} evaluated by the GC protocol.
The process of skipping gates from $f(a, b, p)$ {\it only} utilizes the public input $p$ which is already known to both parties.
In the process, the private values are treated as unknown Boolean variables.
In other words, Alice and Bob do not access their inputs in the \sys{} algorithm for reducing $f(a,b,p)$ to $f_p(a,b)$.
Thus, no information about the private inputs $a$ and $b$ is accessed/revealed by the \sys{} algorithm.
The garbling/evaluation of the two-input Boolean function of $f_p(a,b)$ is passed to the original GC protocol.
Therefore, the security proof of \sys{} is identical to that of the GC protocol.


\section{\frwk{}}\label{sec:arm}
In this section, we present \frwk{}, a GC framework based on a garbled ARM processor and the \sys{} algorithm.
The framework aims to simplify the development of privacy-preserving applications while keeping the garbling cost as low as the best optimized garbled circuits.
We first describe the overview of \frwk{} and its API for GC development.
Then, we explain how ARM's unique architecture helps to decrease garbling overhead.
Next, the effect of \sys{} in reducing the garbling cost is discussed.
Finally, we discuss why we do not employ Oblivious RAM for ARM register files.

\subsection{Global Flow}\label{ssec:arm-global}
The \frwk{} framework allows users to write a two-party SFE program in C/C++ (or any language that can be compiled to the ARM binary code).
\fig{fig:frwk_overview} shows the overview of the framework.
The SFE program is compiled using an ARM cross-compiler, e.g., \texttt{gcc-arm-linux-gnueabi}.
The compiled binary code is fed to the \sys{} algorithm as the public input $p$.
The Boolean circuit that is going to be garbled/evaluated is the synthesized ARM processor circuit.
The \frwk{} framework supports the following API:
\begin{lstlisting}[language=C,basicstyle=\ttfamily,keywordstyle=\color{blue}\ttfamily,stringstyle=\color{red}\ttfamily,commentstyle=\color{CommentColor}\ttfamily]
void gc_main(
  const int *a,// Alice's input
  const int *b,// Bob's input
  int *c) {// output array
  // The user's code goes here.
}
\end{lstlisting}

The entry function, \texttt{gc\_main}, receives three arguments: pointers to Alice's input, Bob's input, and the output.
The framework has five separate memory elements (consisting of flip-flops and MUXs) to store: Alice's inputs, Bob's inputs, output, stack, and instructions.
The flip-flops in the instruction memory are initialized with the compiled binary code that is known to both parties (the public input $p$).
The flip-flops in Alice's and Bob's memories are initialized with the labels corresponding to their private inputs $a$ and $b$, respectively.
The other flip-flops in the stack, output, pipeline registers, and the register file are initialized to zero.
The ARM circuit is garbled following the sequential garbling process~\cite{songhori2015tinygarble} for a pre-specified number of clock cycles.

\begin{figure}[ht]
\centering
\includegraphics[width=0.45\textwidth]{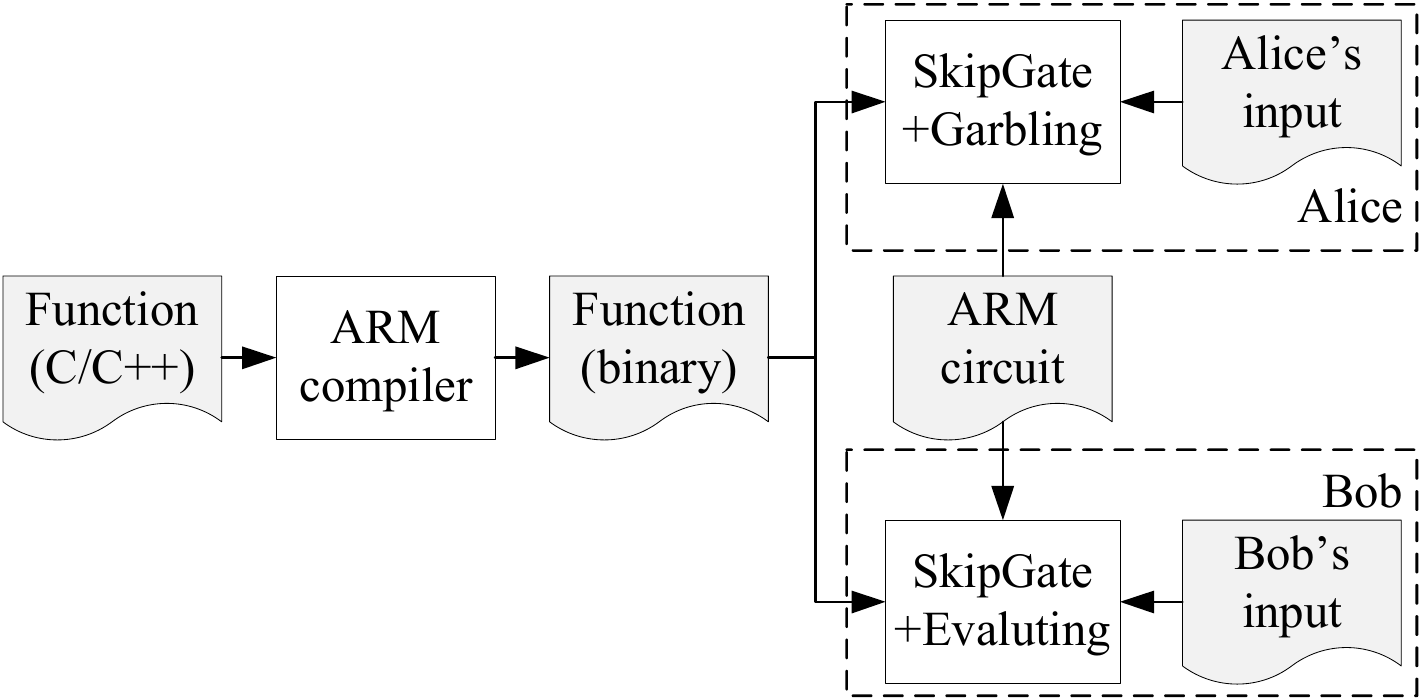}
\caption{Overview of the \frwk{} framework.}\label{fig:frwk_overview}
\end{figure}


\subsection{ARM as a Garbled Processor}\label{ssec:arm}
In this work, we choose ARM as our garbled processor which is a more ubiquitous and sophisticated processor compared to MIPS~\cite{songhori2015tinygarble, wang2015secure, songhori2016garbledcpu}.
ARM has two main advantages:
(1) Pervasiveness: the compilers and toolsets of ARM are under constant scrutiny, updating, and probably, more optimized as a result.
(2) Conditional Execution: Designed to improve performance and code density, conditional execution in ARM allows each instruction to be executed only if a specific condition is satisfied~\cite{sloss2004arm}.

ARM compilers tend to replace conditional branches with conditional instructions to make the flow of the program predictable, and thus, lower the cost of branch misprediction.
Similarly, in a garbled processor, the main design effort is to make sure that the flow of the program is predictable so that the next instruction remains public.
Replacing conditional branches with conditional instructions in garbled ARM generates a code with a predictable flow.
\fig{fig:conditional_exec} shows an example function compiled into assembly with and without the conditional execution.
For the code without the conditional execution, the program counter becomes dependent on the results of the comparison. If one of the compared values are secret, the program counter becomes secret as well. For code with the conditional execution, the program counter goes serially through all the commands serially, irrespective of the result of the comparison operation. Thus, it always remains public. 
We also modify the ARM controller such that conditional instructions always take the same number of cycles regardless of their condition (taken or not taken). Otherwise, the program flow will be dependent on the secret condition. 
Having a secret program counter makes the \sys{} algorithm less effective on \frwk{} and therefore reduces the efficiency of the execution.

\begin{figure}[t]
    \centering
    \begin{subfigure}{0.40\columnwidth}
        \centering
        \includegraphics[width=\textwidth]{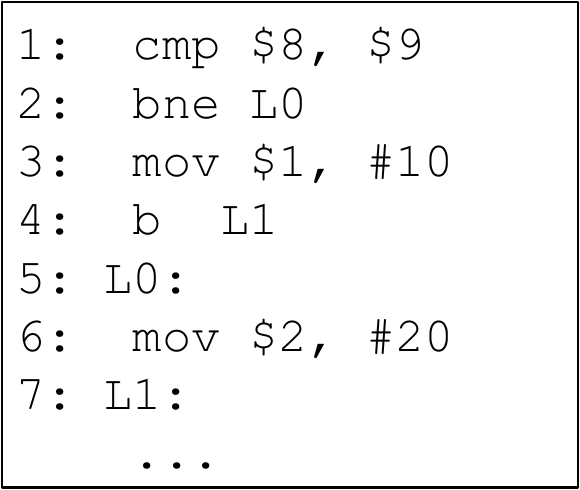}
        \caption{Without Conditional Execution}
    \end{subfigure}
    ~
    \begin{subfigure}{0.40\columnwidth}
        \centering
        \includegraphics[width=\textwidth]{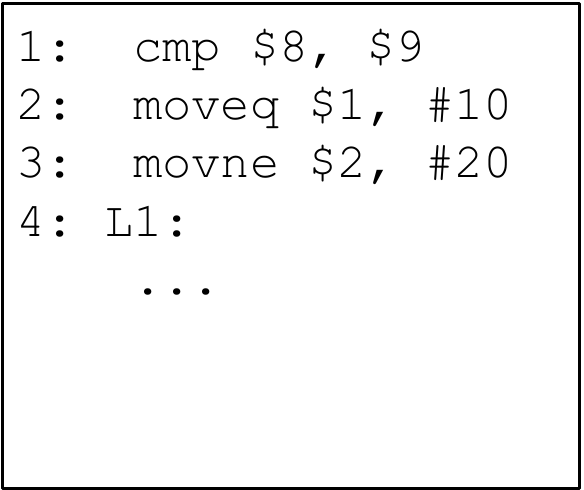}
        \caption{With Conditional Execution}
    \end{subfigure}
    \caption{An example illustrating how conditional execution in ARM can reduce the code size and make the program flow predictable.}\label{fig:conditional_exec}
\end{figure}

We modify and remove a few features from the ARM processor such as interrupts, co-processors, and performance-related components including cache and pipeline. 
These components do not bring any performance advantages in the GC protocol, as the circuit is garbled/evaluated gate by gate (serially).
Note that unlike in hardware, the performance of GC does not increase by parallelizing the gates in the circuit.
In the GC protocol, the total number of garbled non-XOR gates is the {\it only} factor affecting the performance.
The user does not need to pass any flag to the ARM compiler because of the removed components since such blocks only enhance the performance of the processor internally. Therefore, the compiled instructions do not have to be modified because of this modification. 

Implementation of the ARM processor results in a complex and large netlist ($\approx 5$ times larger than that of a MIPS processor).
Thus, using ARM instead of MIPS in the earlier garbled processor approaches~\cite{wang2015secure, songhori2016garbledcpu} would incur an even higher cost.
However, the majority of the components of the ARM processor remain idle during execution of an instruction.
In the next section, we describe how \sys{} utilizes this characteristic to minimize the cost of garbling the processor.

\subsection{Effect of \sys{} on \frwk{}}
As explained above, the instruction memory of the ARM processor is initialized with public values (compiled program). Therefore, if the program counter (the address of the next instruction) is public, the content of the next instruction becomes public as well.
As a result, the control path also becomes public and \sys{} can easily detect the idle components to mark them for skipping.
Moreover, due to \sys{}, the gates of the active components that are only transporting data between memory, register file, and ALU act as wires and do not incur any cost.
According to \sys{}'s notation, the ARM Boolean circuit is a 3-input function $c = f(a,b,p)$ where $p$ is the public binary code and $a$ and $b$ are the parties' private inputs.
\sys{} reduces the ARM circuit into a smaller circuit of $c = f_p(a,b)$ where $f_p$ is able to perform the exact operation required by the public binary code $p$, e.g., $c = a + b$.
Therefore, the main garbling cost is paid only for the actual computation on the secret values.
As explained in the previous section, \sys{} performs these optimizations at the gate level, in contrast to instruction level as in~\cite{wang2015secure, songhori2016garbledcpu}.

\subsection{Why not Sub-linear Oblivious RAM?}
As mentioned in \sect{ssec:arm-global}, we use an array of MUXs and flip-flops to implement the register file in the ARM circuit.
This means that the cost of accessing the register file, when performed obliviously, is linear with respect to its size.
One natural question would be why we did not employ Oblivious RAM (ORAM) that enables oblivious access to memories in the GC protocol with sub-linear cost~\cite{wang2014scoram, zahur2016revisit}.
The reason is that, in most cases, the access to the register file is not required to be oblivious.
Since the instructions come from a publicly known instruction memory, both parties know which register is accessed.
The \sys{} algorithm utilizes this information to skip garbling of the gates in the MUXs of the register file, thus, no garbling cost is required for such accesses.
With ORAM, all the accesses to the register file would be the costly oblivious access.

In rare occasions where two or more instructions should be garbled at a time, accessing a register would not be free using MUXs.
These cases only happen when ARM compiler fails to replace a conditional branch on a secret value with conditional instructions.
The user can typically alter the program in a way that the compiler avoids such branches and replaces it with conditional instructions instead.
However, in these cases, the \sys{} algorithm removes most of the gates in the register file.
Currently, state-of-the-art ORAM constructions such as Circuit ORAM~\cite{wang2015circuit}, SR-ORAM~\cite{zahur2016revisit}, or Floram~\cite{doerner2017scaling} start outperforming the linear scan (MUXs and flip-flops) from memory size of  
8KB (512-bit block size),
8KB (32-bit block size),
2KB (32-bit block size), respectively. 
ARM's total register file has 16 registers, each containing a 32-bit value, thus, the total size of the register file is 64B which is smaller than the break-even points of ORAMs.

\begin{figure}[h]
\centering
\includegraphics[width=0.60\columnwidth]{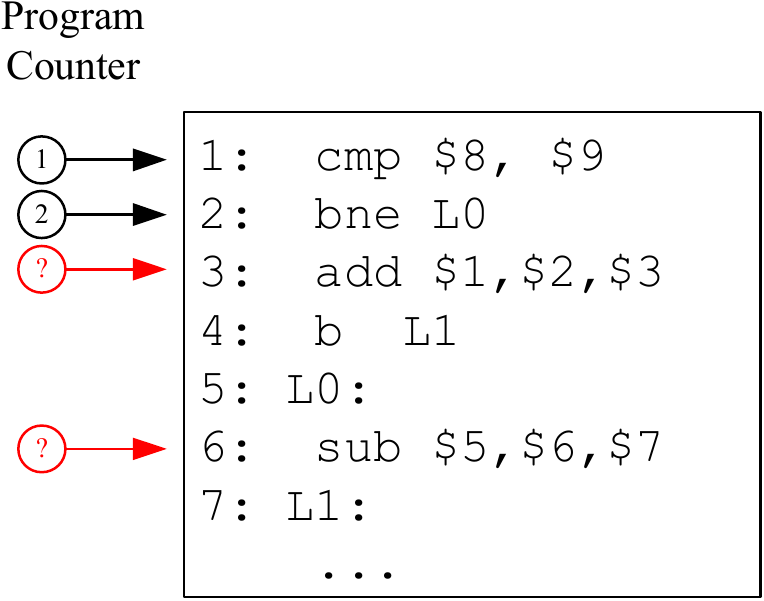}
\caption{Failure to replace a secret branch with conditional instructions, makes the program counter secret.
Thus, the instruction becomes secret.}
\label{fig:branch}
\vspace{-5pt}
\end{figure}

\fig{fig:branch} shows an example where after execution of a branch on a secret value, the next instruction becomes secret and unknown to parties.
In this example, the program counter can either be 3 or 6 depending on the outcome of the comparison in Line 1.
Thus, two instructions \texttt{add \$1, \$2, \$3} (\texttt{\$3 = \$1 + \$2}) and \texttt{sub \$5, \$6, \$7} (\texttt{\$5 = \$6 - \$7}) have to be garbled/evaluated at the same time.
For fetching the second instruction from the register file, we only have two choices: \texttt{\$2} and \texttt{\$6}.
This means that, instead of having a complete oblivious access to the register file with 16 choices, we only have to obliviously select between 2 of the 16 registers.
This costs far less than using ORAMs.
The cost of oblivious access using MUXs and \sys{} to a {\it subset} of a memory is equal to an oblivious access to a memory with the size of the subset.

%

The rationale for using an array of MUXs in the register file also applies to the code, data, and stack memories where the access is almost always public and known to both parties.
In the worst case, only a subset of memory is accessed obliviously, thus making the cost of memory access below the threshold of switching to ORAMs.
The integration of the \sys{} algorithm and garbled processor introduces an unusual use case for oblivious memory where oblivious access is performed only on a varying subset of the memory.
The subset can be different from one access to the other.
The current sub-linear ORAM protocols cannot address this scenario efficiently. Thus, an interesting research question is raised:

\textbf{Is it possible to \textit{obliviously} access (read/write) a varying subset of the memory with a \textit{sub-linear} cost in terms of the subset size?}

Note that programs which contain secret terminate conditions in the for-loops prohibit the protocol to be terminated. The underlying reason is that when the number of loop executions is secret, it is not known when the program is going to finish. This is, in fact, a fundamental constraint for any high-level secure computation frameworks and it is not a shortcoming of \frwk{}. 
By definition, securely computing for-loops for {\it secret} number of times, requires that the protocol's behavior is not dependent on the value of the secret condition. Since the value of the secret condition can be arbitrarily large, the protocol should not terminate until after performing dummy operations for the maximum possible number of loop executions which is prohibitively expensive. 
\section{Evaluation}\label{sec:eval}
\subsection{Evaluation Setup}
We use Synopsis Design Compiler (DC) H-2013.03-SP4~\cite{tool:DesignCompiler} along with TinyGarble~\cite{songhori2015tinygarble} synthesis and technology libraries to generate the netlists for the benchmark circuits and the ARM processor.

For the \frwk{} framework, we use the Amber ARM project, an open-source implementation of ARM v2a ISA on opencores~\cite{santifort2010amber}.
The ARM circuit is modified as explained in \sect{ssec:arm}.
Synthesizing the ARM processor with Synopsis DC takes a few hours.
However, the process is done only once for a given memory size and it can be used for any set of functions and inputs afterward.
The benchmark functions for \frwk{} are implemented in C and compiled using GNU \texttt{gcc-arm-linux-gnueabi} (Ubuntu/Linaro 5.3.1-14ubuntu2).
We used \texttt{-Os} compiler optimization flag in order to reduce the number of instructions.
We modified the header assembly code to change the addresses of stack, code, and data memories in the compiled binary.
We do not apply any optimization on the binary code.
Thus, similar to a normal software compilation, it takes less than a second to compile the majority of the functions into the ARM binary codes.

\subsection{Benchmark Functions and Metrics}
We use the benchmark functions that have frequently been used for evaluation in the GC literature~\cite{buscher2017compiling, mood2016frigate,  songhori2015tinygarble}. 
The most important metric to compare the cost of garbling is the total number of garbled non-XOR gates.
This metric encompasses both the cost of computation (encrypting/decrypting garbled tables) and the cost of communication (transferring garbled tables) in the GC protocol due to the free-XOR optimization~\cite{kolesnikov2008improved}.

\subsection{Effect of \sys{} on Sequential GC}
As described in \sect{sec:skipgate}, the \sys{} algorithm avoids redundant garbling/evaluation of gates in sequential circuits with public wires.
In the sequential benchmark circuits reported in TinyGarble~\cite{songhori2015tinygarble}, the flip-flops were initialized with known values but their output wires were treated as secret.
We applied \sys{} to the same benchmark functions to demonstrate the cost reduction even for a small number of public values.
In \tab{tab:sys_impvoment}, we compare the cost of garbling for circuits generated by TinyGarble~\cite{songhori2015tinygarble} with and without applying the \sys{} algorithm.
As can be seen, cost reduction of \sys{} can be as high as $59.5\%$ for AES and as little as $0\%$ in Compare function.

\begin{table}[t]
\centering
\caption{Improvement by the \sys{} algorithm on sequential circuits generated by TinyGarble~\cite{songhori2015tinygarble}. These functions do not have public inputs. \sys{} benefits from public initial values of the small number of flip-flops to reduce their garbling cost.}\label{tab:sys_impvoment}
\scalebox{0.85}{%
    \begin{tabular}{l||r|r|r||r}
       \multirow{2}{*}{\textbf{Function}}   & \multicolumn{2}{c|}{\textbf{\# of Garbled Non-XOR}} & \multicolumn{1}{c||}{\textbf{\# of}}      & \multirow{2}{*}{{\textbf{Improv.}}}  \\\cline{2-3}
     & \multicolumn{1}{l|}{\textbf{w/o \sys{}}} & \multicolumn{1}{l|}{\textbf{w/ \sys{}}} & \multicolumn{1}{c||}{\textbf{Skipped}} & \bigstrut[b]\\
    \hline
    \hline
    \textbf{Sum 32} & 32    & 31    & 1     & 3.13\% \bigstrut[t]\\
    \textbf{Sum 1024} & 1,024 & 1,023 & 1     & 0.10\% \\
    \textbf{Compare 32} & 32    & 32    & 0     & 0.00\% \\
    \textbf{Compare 16,384} & 16,384 & 16,384 & 0     & 0.00\% \\
    \textbf{Hamming 32} & 160   & 145   & 15    & 9.38\% \\
    \textbf{Hamming 160} & 1,120 & 1,092 & 28    & 2.50\% \\
    \textbf{Hamming 512} & 4,608 & 4,563 & 45    & 0.98\% \\
    \textbf{Mult 32} & 2,048 & 2,016 & 32    & 1.56\% \\
    \textbf{MatrixMult3x3 32} & 25,947 & 25,668 & 279   & 1.08\% \\
    \textbf{MatrixMult5x5 32} & 120,125 & 119,350 & 775   & 0.65\% \\
    \textbf{MatrixMult8x8 32} & 492,032 & 490,048 & 1,984 & 0.40\% \\
    \textbf{SHA3 256} & 40,032 & 38,400 & 1,632 & 4.08\% \\
    \boldmath{}\textbf{AES 128$^\dagger$}\unboldmath{} & 15,807 & 6,400 & 9,407 & 59.51\% \\
    \end{tabular}%
}~\\
\footnotesize{{$\dagger$}The missing key expansion module to AES 128 of~\cite{songhori2015tinygarble} is added here.}
\end{table}

The degree of improvement depends on the structure of the circuit and whether or not the registers are connected to non-XOR gates.
For example, in AES, garbling of the controller part of the sequential circuit (including a counter keeping track of the AES round and MUXs connecting to it) is avoided by \sys{} because both parties know the AES control path in advance.
Note that the functions in \tab{tab:sys_impvoment} do not have any public known inputs that are the main target of \sys{}.
Nevertheless, \sys{} reduces the cost of GC by leveraging the public initial value of the small number of flip-flops in the circuits.

\vspace{0.1in}
{\bf Comparison with Garbled MIPS~\cite{wang2015secure}.}
Even though the approach of \frwk{} is similar to the garbled MIPS presented in~\cite{wang2015secure}, it outperforms that work by a long margin. For example, to compute the Hamming distance between 32 32-bit integers \footnote{as reported in~\cite{wang2015secure}, this is different from the common approach of computing Hamming distance where the inputs are binary},~\cite{wang2015secure} needs 481K garbled gates, whereas \frwk{} needs only 3073- and improvement by 156$\times$.

\subsection{\frwk{} vs HDL Synthesis}
\tab{table:hw_vs_frwk} compares the cost of garbling of (i) functions devised in Verilog HDL and constructed by the hardware synthesis technique of TinyGarble~\cite{songhori2015tinygarble} with (ii) functions developed in C and constructed by the \frwk{} framework.
\sys{} is applied in both cases. 
As expected, \frwk{} incurs only a small overhead (at most 6.6\% for MatrixMult8x8) compared to hardware synthesis method.
In the case of Hamming distance function, \frwk{} results in even less number of non-XOR gates (up to $77.8\%$ improvement).
Note that we use an efficient binary tree-based method~\cite{huang2011faster} for Hamming distance realization in C.

\begin{table}[t]
\centering
\caption{Comparison of the number of garbled non-XOR gates of \frwk{} with the HDL synthesis approach of TinyGarble~\cite{songhori2015tinygarble}. Both frameworks benefit from \sys{}.}\label{table:hw_vs_frwk}
\scalebox{0.85}{%

    \begin{tabular}{l||r|r||r}
    \multirow{3}{*}{\textbf{Function}}       & \multicolumn{2}{c||}{\textbf{\# of Garbled Non-XOR}} & \multirow{3}{*}{\textbf{Overhead}} \bigstrut[b]\\
\cline{2-3}          & \multicolumn{1}{c|}{\textbf{TinyGarble~\cite{songhori2015tinygarble}}} & \multicolumn{1}{c||}{\textbf{\frwk{}}} &  \bigstrut[t]\\
     & \multicolumn{1}{c|}{\textbf{(Verilog)}} & \multicolumn{1}{c||}{\textbf{(C)}} &  \bigstrut[b]\\
    \hline
    \hline
    \textbf{Sum 32} & 31    & 31    & 0.00\% \bigstrut[t]\\
    \textbf{Sum 1024} & 1,023 & 1,023 & 0.00\% \\
    \textbf{Compare 32} & 32    & 32    & 0.00\% \\
    \textbf{Compare 16,384} & 16,384 & 16,384 & 0.00\% \\
    \textbf{Hamming 32} & 145   & 57    & -60.69\% \\
    \textbf{Hamming 160} & 1,092 & 247   & -77.38\% \\
    \textbf{Hamming 512} & 4,563 & 1,012 & -77.82\% \\
    \textbf{Mult 32} & 2,016 & 993   & -50.74\% \\
    \textbf{MatrixMult3x3 32} & 25,668 & 27,369 & 6.63\% \\
    \textbf{MatrixMult5x5 32} & 119,350 & 127,225 & 6.60\% \\
    \textbf{MatrixMult8x8 32} & 490,048 & 522,304 & 6.58\% \\
    \textbf{SHA3 256} & 38,400 & 37,760 & -1.67\% \\
    \boldmath{}\textbf{AES 128$^\dagger$}\unboldmath{} & 6,400 & 6,400 & 0.00\% \\
    \end{tabular}%
}\\
\footnotesize{{$\dagger$}The missing key expansion module to AES 128 of~\cite{songhori2015tinygarble} is added here.}
\end{table}

\subsection{\frwk{} vs GC Frameworks Supporting High-level Languages}
\tab{table:other_vs_frwk} reports the cost of garbling for the benchmark functions constructed  by \frwk{} and the prior-art GC frameworks Frigate~\cite{mood2016frigate} and CBMC-GC~\cite{buscher2017compiling}. The last column compares \frwk{} with the best of these two. 
In all cases, \frwk{} is either equal or better than the earlier frameworks in terms of garbling cost.
It shows significant improvements in hamming distance and AES, 44$\%$ and 38$\%$ respectively. 
Moreover, as shown in the table, software-level optimizations such as $a = a\ \&\ a$ are automatically performed by the ARM compiler. Such operations can result in compile time or runtime errors in several state-of-the-art frameworks as reported in~\cite{mood2016frigate}. 
Note that we choose to compare with these two frameworks as they outperform the earlier frameworks like Obliv-C~\cite{zahur2015obliv} and OblivM~\cite{liu2015oblivm}. 
Even though the approach of \frwk{} is similar to the garbled MIPS presented in~\cite{wang2015secure}, it outperforms that work by a long margin. For example \frwk{} requires $8.4E3$ and $49E3$ times less garbled non-XOR gates for computing Hamming distances with 32 and 512-bit inputs, respectively.




\begin{table}[t]
\centering
\caption{Comparison of the number of garbled non-XOR gates of \frwk{} with the best prior art solution supporting high-level languages. We choose CBMC-GC and Frigate for comparison as they outperform previous frameworks for these benchmarks. The improvement is shown w.r.t. the best of these two.}\label{table:other_vs_frwk}
\scalebox{0.85}{
    \begin{tabular}{l||r|r|r||r}
           \multirow{2}{*}{\textbf{Function}}  & \multicolumn{3}{c||}{\textbf{Number of non-XORs}} &  \multirow{2}{*}{\textbf{Improv.}}\\
\cline{2-4}  & \multicolumn{1}{c|}{\textbf{\hspace{-1em} CBMC-GC~\cite{franz2014cbmc}}} & \multicolumn{1}{c|}{\textbf{Frigate~\cite{mood2016frigate}}} & \multicolumn{1}{c||}{\textbf{\frwk{}}} &  \bigstrut\\
    \hline
    \hline
    \textbf{Sum 32} & -     & \textbf{31} & 31    & 0.00\% \bigstrut[t]\\
    \textbf{Sum 1024} & -     & \textbf{1,025} & 1,023 & 0.20\% \\
    \textbf{Compare 32} & -     & \textbf{32} & 32    & 0.00\% \\
    \textbf{Compare 16,384} & -     & \textbf{16,386} & 16,384 & 0.01\% \\
    \textbf{Hamming 160} & \textbf{449} & 719   & 247   & 44.99\% \\
    \textbf{Mult 32} & -     & \textbf{995} & 993   & 0.20\% \\
    \textbf{MatrixMult5x5 32} & \textbf{127,225} & 128,252 & 127,225 & 0.00\% \\
    \textbf{MatrixMult8x8 32} & \textbf{522,304} & -     & 522,304 & 0.00\% \\
    \textbf{AES 128} & -     & \textbf{10,383} & 6,400 & 38.36\% \\
    \textbf{a = a \texttt{op}$^\ddagger$ a} & \textbf{0} & \textbf{0} & 0     & 0.00\% \\
    \textbf{SHA3 256} & -     & -     & 37,760 & - \\
    \end{tabular}%
}
\footnotesize{{$\ddagger$}\texttt{op} represents any Boolean operation ($+,\&, \oplus,\ etc.$)}
\end{table}

\subsection{Effect of \sys{} on ARM}
\tab{tab:sys_improvment_frwk} shows the cost of garbling an ARM processor for the benchmark functions using conventional GC compared to GC with the \sys{} algorithm.
Since the instruction memory is known to both parties in ARM, \sys{} omits a significant number of non-XOR gates in the circuits.
The circuit of ARM has 126,755 non-XOR gates and for computing a function, for example, Hamming 160, it takes 1,909 clock cycles.
It means with the conventional GC protocol, garbling/evaluation of $1,909\times126,755=241,975,295$ non-XORs is required.
\sys{} reduces the circuit into a smaller circuit with only 247 non-XORs (almost {\it seven orders of magnitude less}).
In the case of AES, we achieve more than {\it six orders of magnitude} improvement over the garbled processor based on the conventional GC without the \sys{} algorithm.
The algorithm transforms the impractical cost of garbling an ARM processor into the near-optimal cost of the reduced circuit.
These dramatic improvements are due to a large number of public inputs in the ARM processor originating from the instruction memory that allows \sys{} to skip garbling/evaluation most of the gates in the ARM circuit.


\begin{table}[t]
\centering
\caption{Improvement by \sys{} on \frwk{}.}
\label{tab:sys_improvment_frwk}
\scalebox{0.85}{%
\begin{tabular}{l||r|r||r}
\multirow{2}{*}{\textbf{Function}} & \multicolumn{2}{c||}{\textbf{\# of Garbled Non-XOR}}  & \multicolumn{1}{c}{\textbf{Improv.}} \\\cline{2-3}
  & \multicolumn{1}{l|}{\textbf{w/o \sys{}}} & \textbf{w/ \sys{}} & \textbf{(1000X)} \Tstrut\\[0.1cm]
  \hline 
Sum 32 & 3,817,680 & 31 & 123 \\
Sum 1024 & 76,483,260 & 1,023 & 75 \\
Compare 32 & 4,072,192 & 130 & 31 \\
Compare 16,384 & 1,047,095,280 & 16,384 & 64 \\
Hamming 32 & 67,063,912 & 57 & 1,177 \\
Hamming 160 & 242,931,704 & 247 & 984 \\
Hamming 512 & 863,559,216 & 1,012 & 853 \\
Mult 32 & 4,199,448 & 993 & 4 \\
MatrixMult3x3 32 & 72,790,432 & 27,369 & 3 \\
MatrixMult5x5 32 & 286,071,488 & 127,225 & 2 \\
MatrixMult8x8 32 & 1,079,894,416 & 522,304 & 2 \\
SHA3 256 & 29,354,783,052 & 37,760 & 777 \\
AES 128 & 54,621,701,856 & 6,400 & 8,535 \\
\end{tabular}
}
\end{table}

\subsection{Complex Functions}
We develop a number of complex functions, as described below, with the \frwk{} framework. 
In each of these functions, the input is XOR-shared between two parties. 
\tab{tab:complex_funct} shows the improvement for these functions by \sys{} over the state-of-the-art GC.

\para{Bubble-Sort}: This function receives a list of 32 32-bit integers, sorts the list using Bubble Sort algorithm, and then writes the sorted list in the output memory.

\para{Merge-Sort}: This function receives a list of 32 32-bit integers, sorts the list using Merge Sort algorithm, and then writes the sorted list in the output memory.

\para{Dijkstra}: This function receives the adjacency matrix of a directed graph with 64 weighted edges (described as a 32-bit integer), finds the shortest path between a source and other nodes using Dijkstra algorithm, and then writes the corresponding distances in the output memory. 

\para{CORDIC}: COordinate Rotation DIgital Computer receives a degree and a 2D vector described as 32-bit fixed-points (2-bit decimal and 30-bit fraction), computes trigonometric, hyperbolic, or exponential functions according to Universal CORDIC algorithm~\cite{volder1959cordic}, and then writes the final 2D vector in the output memory.
The output vector in CORDIC algorithm converges one bit per iteration; thus, it requires 32 iterations in our case.
The addition, shift, and non-oblivious table lookup are the only required operations in this algorithm.
Universal CORDIC has two modes for updating vector: rotational and vectoring and three modes for lookup table: circular, linear, and hyperbolic.
Combining these two modes allows the user to compute trigonometric, hyperbolic, exponential, square root, multiplication, or division functions.
Among these functions, square root and division have previously been reported in~\cite{hussain2016privacy} and require $12,733$ and $12,546$ non-XOR gates respectively, almost three times more than \frwk{}.

\begin{table}[h]
\centering
\caption{Improvement by \sys{} on \frwk{} for the complex functions.}
\label{tab:complex_funct}
\scalebox{0.85}{%
\begin{tabular}{l||r|r||r}
\multirow{2}{*}{Function (bit)} & \multicolumn{2}{c||}{\# of Garbled Non-XOR}  & \multicolumn{1}{c}{Improv.} \\\cline{2-3}
  & \multicolumn{1}{l|}{\textbf{w/o \sys{}}} & \textbf{w/ \sys{}} & (1000X) \Tstrut\\[0.1cm]
  \hline 
Bubble-Sort32 32 & 1,366,390,620 & 65,472 & 21 \\
Merge-Sort32 32 & 981,712,458 & 540,645 & 2 \\
Dijkstra64 32 & 1,493,339,886 & 59,282 & 25 \\
CORDIC 32 & 228,847,596 & 4,601 & 50 \\
\end{tabular}
}
\end{table}
\section{Related Work} \label{sec:related}
The idea of designing a custom programming language to describe and efficiently compile functions for secure evaluation dates back to Fairplay, the first GC compiler~\cite{malkhi2004fairplay}.
Fairplay introduces a custom language, namely, the Secure Function Definition Language (SFDL).
SFDL compiles to Secure Hardware Description language (SHDL).
More powerful languages and compilers were later presented~\cite{HKSSW10, kreuter2012billion, rastogi2014wysteria}.
The introduction of a custom programming language is neither user-friendly nor versatile when compared with conventional programming languages like C.

Another approach adopted in FastGC~\cite{huang2011faster, HS13}, VMCRYPT~\cite{malka2011vmcrypt}, and ABY~\cite{demmler15aby} for GC circuit generation is to design a library containing implementations of GC optimized sub-circuits in a general-purpose high-level language like Java.
This method requires the user to have a thorough understanding of the circuit description of the secure function as the circuits and their decomposition into sub-circuits has to be specified manually.

The first GC implementation supporting a general purpose language is CBMC-GC~\cite{holzer2012secure} which supports ANSI-C.
However, it supports only a subset of ANSI-C that is not compatible with many important primitives, and therefore, not compatible with legacy code. 
The main drawback of~\cite{holzer2012secure} is the compile-time loop unrolling that makes it scale poorly with the input size.
To cope with this problem, the compiler presented in~\cite{kreuter2013pcf} introduces loops that are specified manually within the code and not unrolled until the GC evaluation. The circuit is stored as a Portable Circuit Format (PCF). 
This compiler supports a more general version of C language.
However, in~\cite{holzer2012secure} and~\cite{kreuter2013pcf}, the code had to be compiled with their custom compiler.
As a result, users cannot benefit from the optimizations provided by general purpose compilers.
Moreover, these compilers are less scrutinized and more prone to bugs.
In contrast, \frwk{} supports any general purpose ARM compiler and thus benefits from all the state-of-the-art optimizations, supports legacy codes, and is fully verified.

The TinyGarble framework~\cite{songhori2015tinygarble} allows a user to describe the function with a Hardware Description Language (HDL) like Verilog or VHDL.
It presents custom GC-optimized libraries which enable synthesis of the HDL code with standard logic synthesis tools, thus, benefiting from the standard hardware optimizations.
TinyGarble also suggests using sequential circuits for GC to solve the scalability issue.
Unlike~\cite{kreuter2013pcf}, it allows to infer loops automatically and to optimize across multiple sub-circuits.
However, TinyGarble limits the programmer to a hardware level language which is less user-friendly than a high-level compiler.
Our work utilizes TinyGarble's methodology to generate the most optimized Boolean circuit for the ARM processor.
The big advantage of \frwk{} is that the function to be evaluated securely can be written in any programming language and compiled with any ARM compiler of choice.

The work in~\cite{wang2015secure} accepts a function as a MIPS machine code, which allows the programmer to describe the function in a language of her choice and compile the function with a standard compiler.
They design a MIPS emulator to securely execute the code.
To avoid emulating a large number of instructions supported by the MIPS machine, they perform a data independent static analysis before execution of the program to build a small instruction bank and ALU circuit tailored for each processor cycle.
In contrast, our approach performs this optimization with bit-precision instead of instruction-precision.
Moreover, this is done in the runtime while the circuit remains the same for each cycle.

To solve the problem of secure conditional branches, Wang et al.~\cite{wang2015secure} propose to pad \texttt{nop} instruction to parallel branches so that their lengths become equal.
This way when the code exits either of the branches, it ends up in the same instruction and the process can continue with less cost.
However, this approach increases the cost for conditional branches.
To mitigate this problem, we propose to use the ARM processor which supports conditional execution and can replace these branches with conditional instructions (see \sect{ssec:arm}).
In rare cases where the ARM compiler fails to replace the conditional branch, we adopted their approach in padding the parallel branches with \texttt{nop} instruction.
Overall, our evaluation shows that \frwk{} outperforms their MIPS framework, for example by 4 orders of magnitudes for Hamming distance function, mostly thanks to the \sys{} algorithm and its bit-precision optimization.

Recently, Mood et al.~\cite{mood2016frigate} performed extensive research on the efficiency and reliability of the current frameworks and found out that most of them suffer from reliability issues. For example, they reported that PAL, KSS, CMBC, Obliv-C, ObliVM, and PCF crashed on programs that should have been compiled correctly. Moreover, KSS, ObliVM, and PCF generated incorrect netlists.
As they discuss in the paper, there are serious limitations for formal verification and due to its impracticality, they limit their analysis to validation by testing. 
This type of testing does not detect all possible flaws in the compilation process.
While many of the issues were later taken care of by the respective developers, this research exposed a serious reliability issue regarding the usage of these compilers.

Frigate~\cite{mood2016frigate} introduces a new C-style language for SFE and the corresponding compiler.
Whereas in our work, we utilize C language with standard ARM cross compiler.
Our work also supports any programming language and its corresponding ARM compiler.
As of now, Frigate only supports three different types (\texttt{uint\_t}, \texttt{int\_t}, and \texttt{struct\_t}).
The user can add her own types but it requires a good understanding of the internal structure of the compiler.
Since these three types have a specific bit length, the final computation is not bit-level efficient.
Frigate divides the program into different functions and creates the circuit by calling the corresponding functions and as a result prohibits the overall circuit optimization.
In contrast, our ARM circuit is optimized globally using state-of-the-art hardware synthesis techniques.
Therefore, our overall platform is based on very well-developed and debugged tools that have been used in industry for many years.
Also, if any new update becomes available for these tools, they can effortlessly be incorporated into our framework.

\begin{table}[t]
\centering
\caption{High-level characteristics of secure computation frameworks, their programming languages, and compilers. ``Cust.'' indicates custom designed-compilers. \textbf{CP: }support for Constant Propagation. \textbf{DCE: }support for Dead Code Elimination. \textbf{DGE: }support for Dynamic Gate Elimination.}
\label{tab:high-level-char}
\scalebox{0.98}{%
\begin{tabular}{llcccc}
Framework  & Lang.                   & Compiler     & CP & DCE & DGE \\\hline
CBMC-GC~\cite{buscher2017compiling}    & ANSI-C                  & Cust.        & yes        & yes             & no               \\
KSS~\cite{kreuter2012billion}        & DSL                     & Cust.        & no         & yes             &    no                  \\
PCF~\cite{kreuter2013pcf}        & ANSI-C                  & Cust.        & yes          & yes             &  no                      \\
ObliVM~\cite{liu2015oblivm}     & DSL                     & Cust.        & no         & no              &    no                  \\
Obliv-C~\cite{zahur2015obliv}     & DSL                     & Cust.        & yes        & yes             &    no                  \\
TinyGarble~\cite{songhori2015tinygarble} & HDL                     & HW Synth. & no         & yes             &  no                    \\
Frigate~\cite{mood2016frigate}    & DSL       & Cust.        & yes        & yes             &     no               \\
\frwk{}        & C/C++$^\dagger$ & ARM  & yes        & yes             &           yes       \\
\end{tabular}}
\footnotesize{{$\dagger$} any language with supported ARM compiler}
\end{table}

It is worth mentioning that \sys{} is different from the ``constant propagation'' and ``dead gate elimination'' techniques introduced in \cite{kreuter2013pcf} and \cite{kreuter2012billion}, respectively. These solutions eliminate parts of the code that do not contribute to the output (or can be computed) at the compile time using static analysis. In contrast, \sys{} performs gate-level optimization dynamically at the runtime to reduce the number of non-XOR gates to close to the optimal value (compared to state-of-the-art HDL frameworks~\cite{songhori2015tinygarble}). Indeed, this is the reason why \frwk{} outperforms \cite{kreuter2013pcf,kreuter2012billion}. For example, for 160-bit Hamming distance, \cite{kreuter2013pcf} reports 880 number of non-XOR gates while \frwk{} garbles only 247. 
\tab{tab:high-level-char} shows a high-level comparison of \frwk{} to the prior secure computation frameworks.

In~\cite{kerschbaum2011automatically} and~\cite{rastogi2013knowledge}, authors address an interesting yet orthogonal problem to ours. They compute what information can be obtained from computation output and each party's private input, whereas, we compute what information can be revealed based on private and public inputs from both parties to avoid garbling/evaluating selected gates. Their approach is inapplicable when only one party is provided with final output and function is required to be evaluated without revealing intermediate values. They do not use a standard verified compiler and cannot garble sequential circuits.

There has been extensive research on secure computation frameworks for machine learning~\cite{mohassel2017secureml,riazi2018chameleon,juvekar2018gazelle,riazisp,riazi2019xonn,riazi2019mpcircuits}. These frameworks are customized for the operations that are frequently performed in the machine learning applications such as vector-dot-product and certain non-linear functionalities. In contrast, the focus of this work is to create a generic high-level secure computation framework.  
\section{Conclusion} \label{sec:conclusion}
This paper introduces the novel \sys{} algorithm for Yao's Garbled Circuit protocol.
The algorithm dynamically omits the communication cost for gates with outputs independent of private data and also the gates not affecting the final output.
Based on the \sys{} algorithm and the ARM processor architecture, we create \frwk{}, a simple-to-use and verified garbled circuit framework.
Users can develop secure functions in high-level languages and compile them using standard ARM cross-compilers.
As a result of \sys{}, only the gates associated with private data in the ARM circuit incur communication and encryption cost.
Evaluations on a host of benchmark functions show that the \frwk{} framework achieves efficiency close to that of HDL-level synthesis methods.


\ifsubmission
  \bibliographystyle{plain}
\else
  \bibliographystyle{alpha}
\fi
\bibliography{0_main}

\end{document}